\documentstyle[multicol,prl,aps,epsfig]{revtex}

\makeatletter

\providecommand{\LyX}{L\kern-.1667em\lower.25em\hbox{Y}\kern-.125emX\@}

\makeatother

\begin{document}

\title{Role of Boron \protect\( p\protect \)-Electrons and Holes in Superconducting
\protect\( MgB_{2}\protect \), and other Diborides: A Fully-relaxed, Full-Potential
Electronic Structure Study}

\author{Prabhakar P. Singh}

\address{Department of Physics, Indian Institute of Technology, Powai, Mumbai -400076,
India}
\date{\today}
\maketitle

\begin{abstract}
We present the results of fully-relaxed, full-potential electronic structure
calculations for the new superconductor \( MgB_{2} \), and \( BeB_{2} \),
\( NaB_{2} \) and \( AlB_{2} \), using
density-functional-based methods. Our results, described in terms of (i) density
of states (DOS), (ii) band-structure, and (iii) the DOS and the charge
density around the Fermi energy \( E_{F} \), clearly
show the importance of \( B \) \( p \)-band for superconductivity.
 In particular, we show that around \( E_{F} \),
the charge density in \( MgB_{2} \), \( BeB_{2} \) and \( NaB_{2} \) is planar
and is associated with the \( B \) plane. For \( BeB_{2} \)
and \( NaB_{2} \), our results indicate qualitative similarities but significant
quantitative differences in their 
electronic structure due to different lattice
constants \( a \) and \( c \). 
\end{abstract}
\begin{multicols} {2}
The intricate interplay of interactions between electrons, holes and phonons
is quite evident in the recently discovered superconductor Magnesium Boride,
\( MgB_{2} \)\cite{akimitsu}. The superconducting temperature of
around \( \approx 39^{0}K \) in \( MgB_{2} \) is almost at the extreme end
of \( T_{c} \)'s predicted by the BCS theory for conventional superconductors.
The ongoing experimental investigations suggest it to be a BCS type 
superconductor
with confirmed isotope effects \cite{budko} and other related effects 
\cite{gray}. However, more experiments
are needed to completely characterize the nature of interaction responsible
for superconductivity (SC) in \( MgB_{2} \). Theoretical studies  of \( MgB_{2} \),
since the announcement of SC by Akimitsu, have concentrated on
(i) the nature of interactions responsible for SC using model
Hamiltonians \cite{hirsch,pickett} and (ii) the electronic structure 
\cite{pickett,kortus,mark,satta} and the linear response 
\cite{pickett,kortus,satta,andersen}
of normal
state \( MgB_{2} \) using \( ab \) \( initio \) methods but with the experimentally
obtained lattice constants \( a \) and \( c \). 

In order to understand SC in \( MgB_{2} \), be it due to heavily
dressed holes \cite{hirsch}, hole-doped covalent bonds \cite{pickett} or simply due to electron-phonon  interaction \cite{osborn},
it is essential to have an accurate electronic structure description of the
normal state \( MgB_{2} \). It is also becoming clear that the boron layer
and the two-dimensional \( \sigma  \) band due to \( p_{x,y} \) orbitals in
that layer hold the key to SC in \( MgB_{2} \) (we present evidence
of that) \cite{pickett,kortus,mark,satta}. Thus an understanding of how the \( \sigma  \) band and the \( p_{x,y} \)
orbitals respond to changes in the chemical environment, lattice constants etc.
could be quite useful in understanding SC in \( MgB_{2} \) and
in synthesizing new materials with \( MgB_{2} \)-like SC. Fortunately,
these changes around \( B \) layer can be mimicked without destroying its two-dimensional
character which is essential for SC. For example, the effects
of removing one electron and adding one electron around the \( B \) layer can
be studied by considering \( NaB_{2} \) and \( AlB_{2} \), respectively. Similarly,
a study of \( BeB_{2} \) (and \( NaB_{2} \)) will provide the effects of changing
the lattice parameters \( c \) and to a lesser extent \( a \). Thus an accurate
and reliable systematic study of the electronic structure of \( MgB_{2} \),
\( BeB_{2} \), \( NaB_{2} \) and \( AlB_{2} \), coupled with the experimental
facts of SC in \( MgB_{2} \) and no SC in \( AlB_{2} \),
can provide substantive clues about the nature of interaction responsible for
SC as well as help in synthesizing new superconducting materials
with \( T_{c} \) 's comparable to that of \( MgB_{2} \). 

As the nature of bonding in boron layer depends critically on lattice constants
\( a \) and \( c \), and since for SC one needs the electronic
structure within a small energy interval of \( \pm hw_{D} \) ( \( \omega _{D}\equiv  \)
Debye frequency) around the Fermi energy, which gets affected the most due to
change in lattice constants, it is of utmost importance to theoretically optimize
the lattice constants \( a \) and \( c \). Some of the earlier studies of
\( MgB_{2} \), which are based on the experimental lattice constants, have
provided a good overall description of the electronic structure 
\cite{pickett,kortus,mark,satta,burdett,ivan}
 and some others
have used the Fermi surface \cite{kortus} to discuss SC in \( MgB_{2} \).
Previous study \cite{mark} of \( BeB_{2} \) was carried out with assumed values for the
lattice constants \( a \) and \( c \). A recent calculation \cite{satta} 
of \( BeB_{2} \)
gives optimized values for \( a \) and \( c \) but we find these values to
be not fully converged. In our opinion, there is a clear need to include (a)
structural relaxation of the lattice, (b) a detailed description of electronic
structure in a small interval around Fermi energy, and (c) a comparison with
other related compounds which have either the potential of becoming an \( MgB_{2} \)-like
superconductor or help in understanding SC in \( MgB_{2} \), 
in a density-functional-based electronic structure calculations of \( MgB_{2} \).
The present work is a step in that direction. 

In the following, we present the results of fully-relaxed, full-potential electronic
structure calculations for \( MgB_{2}, \) \( BeB_{2} \), \( NaB_{2} \) and
\( AlB_{2} \) in \( P6/mmm \) crystal structure, using density-functional-based
methods. We analyze our results in terms of (i) density of states (DOS), (ii)
band-structure along symmetry directions, and (iii) the DOS and the electronic
charge density in a small energy window around the Fermi energy, \( E_{F} \), 
in these systems. 

For our calculations, we have used ABINIT code \cite{abinit}, 
based on psuedopotentials and
planewaves, to optimize the lattice constants \( a \) and \( c \) of \( MgB_{2}, \)
\( BeB_{2} \), \( NaB_{2} \) and \( AlB_{2} \) . In order to study the band-structure
along symmetry directions and site- and symmetry-decomposed densities of states,
we have calculated, self-consistently, the electronic structure of these compounds
using our own full-potential program as well as LMTART package 
\cite{lmtart} with the optimized
lattice constants. For studying the charge density in a small energy window
around \( E_{F} \) we have used the Stuttgart TB LMTO package 
\cite{tblmto}. 

Based on our calculations, described below, we find that in \( MgB_{2} \) (i)
\( B \) \( p \)-band is crucial for SC due to its proximity
to \( E_{F} \). In particular, it is evident that within a small energy window
around \( E_{F} \), the charge density is two-dimensional and it is associated
with the \( B \) plane, (ii) structural relaxation leads to significant changes
around \( E_{F} \) in the densities of states and charge density. For the hypothetical
compounds \( BeB_{2} \) and \( NaB_{2} \), our results indicate qualitative
similarities but significant quantitative differences in electronic structure
due to very different lattice constants \( a \) and \( c \) of these compounds. 
A comparison of
band-structure of \( BeB_{2} \), \( NaB_{2} \), \( MgB_{2} \) and \( AlB_{2} \)
along symmetry directions reveals that the \( \Gamma _{5} \) point moves down,
starting from above \( E_{F} \) in \( NaB_{2} \) to well below \( E_{F} \)
in \( AlB_{2} \). If SC is due to holes then one would expect
\( BeB_{2} \) and \( NaB_{2} \), if they can be synthesized, to become superconducting
but not \( AlB_{2} \). 

Before describing our results in detail, we provide some of the computational
details of our calculations. The structural relaxation was carried out by the
molecular dynamics program ABINIT with Broyden-Fletcher-Goldfarb-Shanno minimization
technique \cite{abinit}  using Troullier-Martins pseudopotentials
\cite{troullier} , 512 Monkhorst-Pack \cite{monkhorst} \( {\bf k} \)-points
and Teter parameterization \cite{abinit} for the exchange-correlation. 
The kinetic energy cutoff
for the plane waves was 110 Ry. The charge self-consistent full-potential LMTO
calculations were carried out using 2$\kappa$-panels with 
the generalized gradient approximation for
exchange-correlation as given by Perdew \emph{et al.} \cite{perdew} 
and 484 \( {\bf k }\)-points in the irreducible
wedge of the Brillouin zone. The basis set used consisted of \( s, \) \( p, \)
\( d \) and \( f \) orbitals at the \( Mg \) site and \( s, \) \( p \)
and \( d \) orbitals at the \( B \) site. The potential and the wavefunctions
were expanded up to \( l_{max}=6 \). The input to the tight-binding LMTO calculations,
carried to charge self-consistency, were similar to that of the full-potential
calculations except for using spherically symmetric potential and the space-filling
atomic spheres. 

In Table I we compare the calculated lattice constants \( a \) and \( c \)
for \( MgB_{2}, \) \( BeB_{2} \), \( NaB_{2} \) and \( AlB_{2} \) with the
experimentally observed values \cite{acta,lipp}
, indicated in the parentheses. The calculated
lattice constants for \( MgB_{2} \) and \( AlB_{2} \) are smaller than their
respective experimental values by roughly \( 1\% \). It compares well with
the differences mentioned in \cite{kortus} for the lattice constants of 
\( MgB_{2} \).
For \( BeB_{2} \) and \( NaB_{2} \) the values obtained for \( c/a \) are
0.98 and 1.43, respectively. Although, one expected the change in \( c/a \)
ratio to be small in \( BeB_{2} \) and large in \( NaB_{2} \) as compared to
\( MgB_{2} \), however, the \( c/a \) value for \( NaB_{2} \) is larger 
than expected. Because of the large difference in the \( c/a \) ratio
of \( BeB_{2} \) and \( NaB_{2} \) they can be used to explore the importance
of inter-layer coupling for SC in these systems. A recent 
calculation \cite{satta}
of \( BeB_{2} \) gives a somewhat smaller value for \( a \) and \( c \).
We have used the lattice constants as given in Ref. \cite{satta} 
 to calculate the total
energy of \( BeB_{2} \) using the full-potential LMTART program, and we find
that the lattice constants as given in Table I of our paper correspond to lower
energy by about \( 3 \) \( mRy \). We have also run the ABINIT code with the
information given (which is not complete) in Ref. \cite{satta} 
to optimize the lattice
constants of \( BeB_{2} \). The results are in agreement with the values given
in Table I. In Table I, we also show the plasma frequencies and the 
total density of states
per spin, \( n(E_{F}) \), at \( E_{F} \) for these compounds. The plasma frequencies
along \( x \) (\( y \) ) direction are very similar for all four diborides
but along the \( z \) direction it correlates with the \( c/a \) ratio of
the compound. A smaller \( c/a \) ratio leads to enhanced coupling in the \( z \)
direction.

We show in Fig. 1 the total density of states for \( BeB_{2} \), \( NaB_{2} \),
\( MgB_{2} \) and \( AlB_{2} \) calculated using the ABINIT program at the
optimized lattice constants of these compounds as given in Table I. The gross
features of the DOS of the four compounds are similar if one takes into account
the differences in the total number of valence electrons. The bottom of the
band is the deepest for \( AlB_{2} \) which has a total of nine valence electrons
and shallowest for \( NaB_{2} \) with only seven valence electrons. However,
the isoelectronic structures \( BeB_{2} \) and \( MgB_{2} \) show substantial
differences in their DOS near the bottom of the band. This is due to the fact
that the smaller \( c/a \) ratio leads to enhanced repulsion which pushes the
\( s \) and \( p \) electrons at the \( Mg \) site downward and, at the same
time, diminishes the DOS for \( B \) \( s \) and \( p \) electrons in the
middle of the band. In the case of \( AlB_{2} \) we find that the \( \textrm{B } \)
\( p \)-band is completely inside \( E_{F} \). 

An important factor in determining the superconducting temperature in conventional
superconductors is the DOS within an interval of \( \pm  \)\( hw_{D} \) at
\( E_{F} \), In order to analyze our results in detail we have plotted in Fig.
2 the symmetry-decomposed DOS at the \( B \) site within a small energy interval
around \( E_{F} \). From the DOS for \( AlB_{2} \), it is evident that it
will not be superconducting, not at least in the same sense as that of \( MgB_{2} \),
because the bands with \( x \) and \( y \) symmetry are completely filled.
According to our calculations as shown in Fig. 2, \( NaB_{2} \) is likely to
show SC with enhanced \( T_{c} \). 

To see how the differences in the lattice 
constants and the number of valence electrons affect the band-structure, 
in Fig. 3 we have
plotted the band-structure along the symmetry directions for 
\( BeB_{2} \),
\( NaB_{2} \), \( MgB_{2} \) and \( AlB_{2} \). 
We find that these differences lead to significant 
changes in the band-structures of 
\( BeB_{2} \),
\( NaB_{2} \), \( MgB_{2} \) and \( AlB_{2} \).  A measure of these 
changes is evident in the movement of
$\Gamma _5$ 
point with respect to $E_F$
as one goes from $BeB_2$ to $AlB_2$.
Due to different number of valence electrons, the $\Gamma _5$ point is 
well above $E_F$ in $NaB_2$ but well below $E_F$ in $AlB_2$.  
The differences between the band-structures of $BeB_2$ and $MgB_2$ arise due 
to differences in the lattice constants. Our calculations show that 
$\Gamma _5$ point is inside $E_F$ by about $5$ $mRy$ in $BeB_2$, while it 
is slightly above $E_F$ in $MgB_2$.  
It is interesting to note that the location of $\Gamma _5$ point, 
around which the Fermi surface in $MgB_2$ 
is hole-like \cite{kortus,armstrong}, has been correlated with SC in 
$MgB_2$ and other diborides \cite{hirsch}.
Our $l$-character analysis shows 
substantial $p_x$ and $p_z$ characters along  $\Gamma$-$A$  direction 
near $E_F$ in all the diborides.

To complete our analysis we have also calculated the electronic charge density
within a \( 5 \) \( mRy \) energy window around \( E_{F} \) for \( BeB_{2} \),
\( NaB_{2} \), \( MgB_{2} \) and \( AlB_{2} \) compounds using the TB LMTO
method with the lattice constants as given in Table I. In these calculations
the atomic sphere radii were adjusted to minimize the discontinuity in the Hartree
potential across the atomic spheres which have been found to give reliable results \cite{kortus,singh}. 
To reduce the overlap errors in \( NaB_{2} \), which has a large \( c/a \)
ratio, we have inserted an empty sphere. In Fig. 4, we show the charge density
calculated within a \( 5 \) \( mRy \) energy window around \( E_{F} \) for
these compounds. For \( MgB_{2} \), \( BeB_{2} \) and \( NaB_{2} \)we find
very similar charge density, which is due to \( p_{x,y} \) orbitals in the
\( B \) plane. It should be noted that the holes in \( MgB_{2} \) also originate
from these planar orbitals in the \( B \) plane. In contrast, the charge density
around \( E_{F} \) in \( AlB_{2} \) has a three-dimensional character and
it consists mostly of \( Mg \) \( s \)- and \( B \) \( p \)-electrons. 

In conclusion, we have presented the results of our fully-relaxed, full-potential
electronic structure calculations for \( BeB_{2} \), \( NaB_{2} \), \( MgB_{2} \)
and \( AlB_{2} \) in \( P6/mmm \) crystal structure using density-functional-based
methods. We have analyzed our results in terms of the density of states, band-structure,
 and the DOS and the electronic charge density 
around $E_F$ in these systems. For \( MgB_{2} \),
we find that the \( p \)-band of \( B \) is crucial for SC
due to its proximity to \( E_{F} \). In particular, we have shown that within
a small energy window around \( E_{F} \), the charge density is two-dimensional
and it is associated with the \( p_{x,y} \) orbitals in the \( B \) plane.
For \( BeB_{2} \) and \( NaB_{2} \), our results indicate qualitative similarities
but significant quantitative differences in the electronic structure due to
differences in the lattice constants. For example,  we find that 
the \( \Gamma _{5} \)
point in \( BeB_{2} \) is just below the Fermi energy 
while in \( NaB_{2} \) 
it is  well above the Fermi energy.

\begin{table}
\caption{The calculated lattice constants $a$ and $c$, the plasmon frequencies,
and the density of states at the Fermi energy. 
The experimental lattice 
constants are shown in the parentheses.}
{\centering \begin{tabular}{|c|c|c|c|c|c|}
\hline 
&
\( a \) (a.u.)&
\( c \) (a.u.)&
\( \omega _{p}^{x} \) (\( eV \))&
\( \omega _{p}^{z} \) (\( eV \))&
\( n(E_{F}) \)\\
\hline 
\hline 
\( BeB_{2} \)&
5.49(5.558)&
5.41(5.425)&
8.1&
11.8&
3.29\\
\hline 
\( NaB_{2} \)&
5.70&
8.15&
7.75&
3.40&
6.75\\
\hline 
\( MgB_{2} \)&
5.76 (5.834)&
6.59 (6.657)&
7.04&
6.77&
4.70\\
\hline 
\( AlB_{2} \)&
5.63 (5.681)&
6.13 (6.149)&
8.11&
10.02&
2.66\\
\hline 
\end{tabular}\par}
\end{table}

\begin{figure}
\centerline{
\epsfig{file=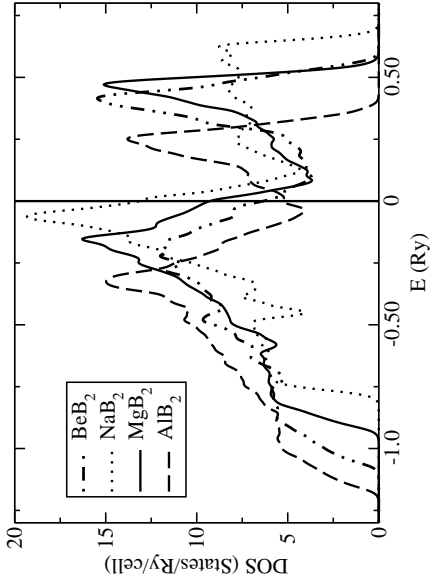,width=0.8\linewidth,clip=true}}
\caption{The total density of states 
calculated at the optimized lattice constants using the ABINIT program.  
     \label{fig.dos}
}
\end{figure}

\begin{figure}
\centerline{
\epsfig{file=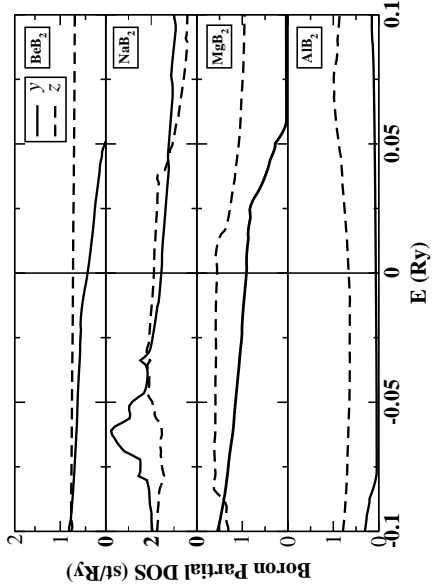,width=0.8\linewidth,clip=true}}
\caption{The partial $B$ density of states around  Fermi energy 
calculated at the optimized lattice constants using the full-potential LMTO 
method.  
     \label{fig.dosef}
}
\end{figure}
\begin{figure}
\centerline{
\epsfig{file=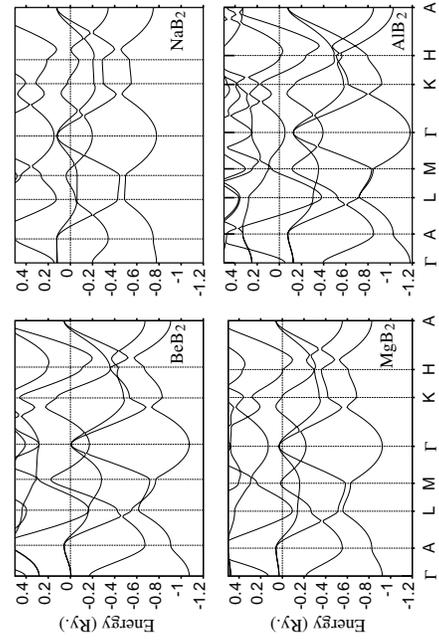,width=0.9\linewidth,clip=true}}
\caption{The band-structure along symmetry directions 
calculated at the optimized lattice constants using the full-potential LMTO 
method.  
      \label{bands}
}
\end{figure}
\begin{figure}
\centerline{
\epsfig{file=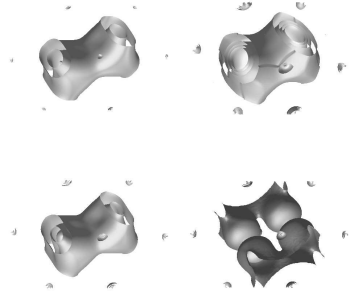,width=0.9\linewidth,clip=true}}
\caption{The isosurfaces of charge density within a $5$ $mRy$ energy window 
around the Fermi energy in the primitive cell of 
$BeB_2$ (top left), $NaB_2$ (top right), $MgB_2$ (bottom left) and $AlB_2$ 
(bottom right),  
calculated at the optimized lattice constants using the TB LMTO 
method. The values of the isosurfaces 
(8,6,4,2,1)
decrease as one 
moves away from the $B$ plane.  The unit for charge density in the volume 
is $10^{-4}[e/(a.u.)^3]$ and
the view is off-diagonal at an angle of $30^0$. 
      \label{charge}
}
\end{figure}
\end {multicols}
\end{document}